\documentclass[a4paper]{VisionStyle}
\usepackage{epsfig}

\begin{document}

\title{Elemental Abundances in Stellar Coronae with XMM-Newton}

\author{M.\,Audard \and M.\,G\"udel} 

\institute{
  Paul Scherrer Institut, W\"urenlingen and Villigen, 5232 Villigen PSI,
  Switzerland}

\maketitle 

\begin{abstract}
The XMM-Newton Reflection Grating Spectrometer Team has obtained observations of
a large number of coronal sources of various activity levels, ages, and spectral
types. In particular, X-ray bright RS CVn binary systems display saturated
coronal emission with spectral lines characteristic of hot (10-30 MK) plasma.
Furthermore, we have obtained XMM-Newton data from young solar analogs 
both within and outside the X-ray saturation reg\-ime. We have simultaneously
analyzed the EPIC MOS and RGS data from these objects and have obtained 
coronal abundances of various elements (e.g., C, N, O, Ne, Mg, Si, Fe).
We show that there is evidence for a transition from an Inverse
First Ionization Potential (FIP) effect in most active stars to a ``normal'' 
solar-like FIP effect in less  active stars. We discuss this result with regard 
to photospheric abundances.

\end{abstract}

\section{Introduction}
High-resolution X-ray spectra of stellar coronae obtained by \textit{XMM-Newton}
and \textit{Chandra} now allow us to study in detail the
rich forest of X-ray lines emitted by elements abundant in stellar coronae, such as
C, N, O, Ne, Mg, Si, S, Ar, Ca, Fe, and Ni. In the past, stellar coronal 
abundances have frequently been determined using the moderate 
spectral resolution of CCD spectra from \textit{ASCA} (e.g.,
\cite{maudard-B1:drake96,maudard-B1:guedel99})
or from the low sensitivity spectrometers onboard \textit{EUVE} (e.g.,
\cite{maudard-B1:drake95,maudard-B1:laming96,maudard-B1:schmitt96,maudard-B1:drake97}). 
The abundance pattern in stellar coronae is complementary to the 
abundance pattern in the Sun: the solar corona, the solar wind, and solar
energetic particles (and probably also galactic cosmic rays) display a so-called ``First Ionization Potential'' (FIP) 
effect, for which abundances of low-FIP ($<10$~eV) elements are enhanced 
relative to their respective photospheric abundances, while the abundances of 
high-FIP ($>10$~eV) elements are photospheric (\cite{maudard-B1:feldman92}, see 
also \cite{maudard-B1:meyer85}). Stellar coronal observations however
often showed a deficiency of metals relative to the solar photospheric 
abundances (\cite{maudard-B1:schmitt96}). \textit{EUVE} spectra either indicated 
the absence of any FIP-related bias (\cite{maudard-B1:drake95}), 
or a solar-like FIP effect (\cite{maudard-B1:drake97}) in inactive stellar coronae. 
The new X-ray observatories \textit{XMM-Newton} and \textit{Chandra} 
combine the high spectral resolution with moderate effective areas to routinely 
obtain data allowing to measure the abundances in stellar coronae.

Recently, \cite*{maudard-B1:brinkman01} showed a trend towards enhanced
high-FIP elemental abundances, while low-FIP
abundances are depleted; this effect was dubbed the ``Inverse FIP'' (IFIP) effect.
Other active stars showed a similar trend
(\cite{maudard-B1:guedel01a,maudard-B1:guedel01b}), except the
intermediately active Capella (\cite{maudard-B1:audard01a}). 
Note however that stellar coronal abundances have often been normalized to the \emph{solar} photospheric abundances, while they should
better be normalized to the \emph{stellar} photospheric abundances. The latter
are difficult to measure. Nevertheless, for some stars, photospheric abundances 
are known. The uncertainty introduced by photospheric  abundances can then be 
removed. We will show that there is a transition from an IFIP to a normal FIP effect in the long-term evolution of the coronae from active to inactive 
solar analogs. We will use data of bright active RS CVn binary systems to complement 
the study as well. Finally, the variation of stellar coronal abundances during
flares will be discussed.

\section{Observations and Data analysis}
\label{maudard-B1_sec:obs}
\textit{XMM-Newton} observed several coronal sources as part of the RGS
stellar Guaranteed Time Program. Observations of active bright RS 
CVn binary systems (e.g., \object{HR1099}, \object{Capella}, \object{UX Ari}, 
\object{$\lambda$ And}) provided excellent high-resolution RGS spectra. The solar
past has also been pro\-bed with observations of young solar analogs of
different ages and activity levels (e.g., \object{AB Dor}, \object{EK Dra},
\object{$\pi^1$ UMa}, \object{$\chi^1$ Ori}). The RGS1, RGS2, and EPIC MOS2
spectra have been simultaneously fitted (except for Capella
where there are no EPIC data available) in XSPEC 11.0.1aj
(\cite{maudard-B1:arnaud96}) using the
\texttt{vapec} model (APEC code with variable abundances). Because of the inaccuracy
and incompleteness of atomic data for non-Fe L-shell transitions, significant
parts of the RGS spectra had to be discarded above 20~\AA. Furthermore, some 
Fe L-shell lines with inaccurate atomic data were not fitted. For additional
information on the data analysis, we refer to \cite*{maudard-B1:audard02} and
\cite*{maudard-B1:guedel02}.

\section{Results}
\label{maudard-B1_sec:res}

\subsection{Coronal abundances of active stars}

\begin{figure*}[!ht]
  \begin{center}
     \includegraphics{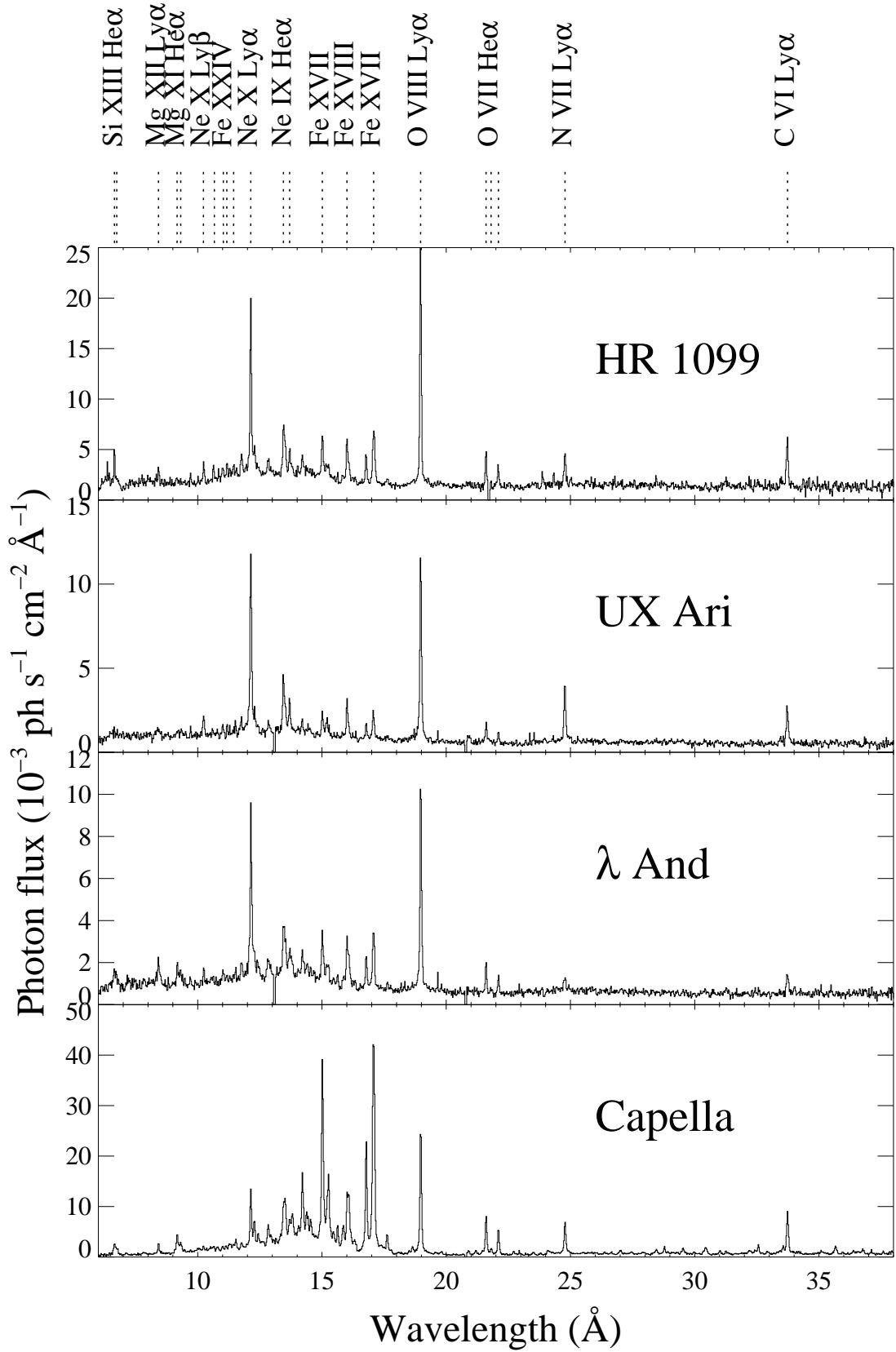}	
  \end{center}
\caption{RGS spectra of bright active RS CVn binary systems. The sources have
been ordered with decreasing activity levels (or average
coronal temperature) from top to bottom. Major emission lines have been labeled.}  
\label{maudard-B1_fig:fig1}
\end{figure*}

\begin{figure*}[!ht]
  \begin{center}
    \includegraphics{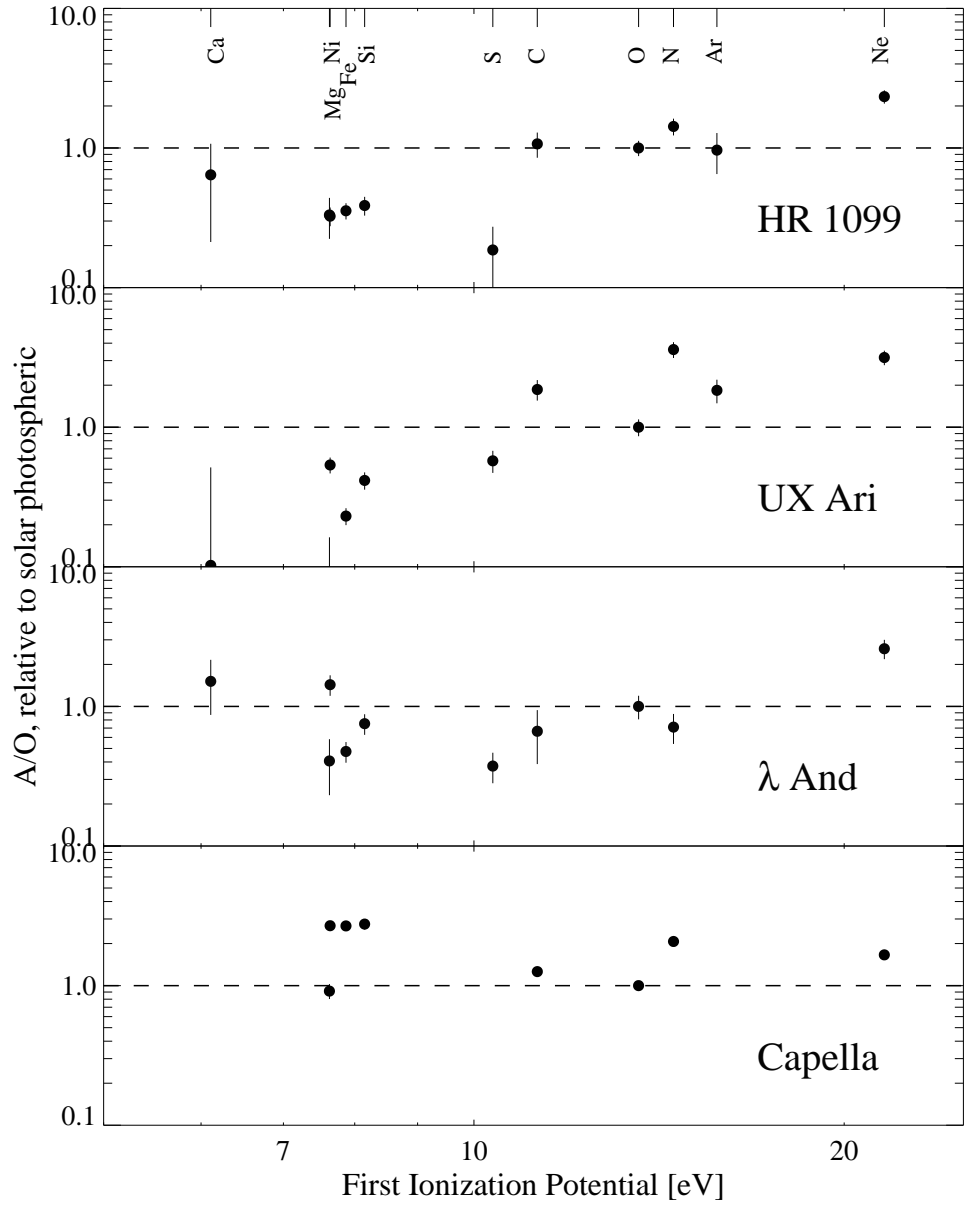}	
  \end{center}
\caption{Coronal abundance normalized to oxygen in RS CVn 
binaries as a function of the First Ionization Potential. Solar photospheric 
abundances from Anders \& Grevesse (1989) were used, except for {\rm Fe} 
(Grevesse \& Sauval 1999).}  
\label{maudard-B1_fig:fig2}
\end{figure*}

The RGS spectra of four RS CVn-type systems shown in Figure~\ref{maudard-B1_fig:fig1}
display bright H-like and He-like emission lines (mostly from C, N, O, Ne, Mg,
Si) and numerous Fe L-shell lines. Although the overall spectral features are
similar in HR~1099, UX Ari, and $\lambda$ And, there are differences in terms
of line intensities or ratios. This can suggest differences in the emission
measure distributions, or in the elemental composition in their coronae. A hint
pointing toward the latter interpretation comes from the different ratios of
the Fe~\textsc{xvii} (at 15\AA) and Ne~\textsc{ix} lines, which have similar
maximum formation temperatures. The line ratio in Capella
definitely points toward an abundance effect, with a high Fe/Ne ratio. We have performed fits
(see \S\ref{maudard-B1_sec:obs}) to the data and modeled the X-ray spectra. Coronal
abundances (normalized to the oxygen abundance to remove the uncertainty that
might be introduced by the determination of the underlying continuum)
relative to the solar photospheric abundances are shown in
Fig.~\ref{maudard-B1_fig:fig2} as a function of the first ionization potential of
the element. A similar trend (IFIP effect) as found by \cite*{maudard-B1:brinkman01} is seen in UX Ari,
with low-FIP elements depleted relatively to the high-FIP elements, and
confirmed in this new analysis of HR~1099. In contrast, the FIP bias in
$\lambda$ And is less clear. The absence of any FIP bias is suggested in the
intermediately active binary Capella.

\begin{figure*}[!ht]
  \begin{center}
    \includegraphics{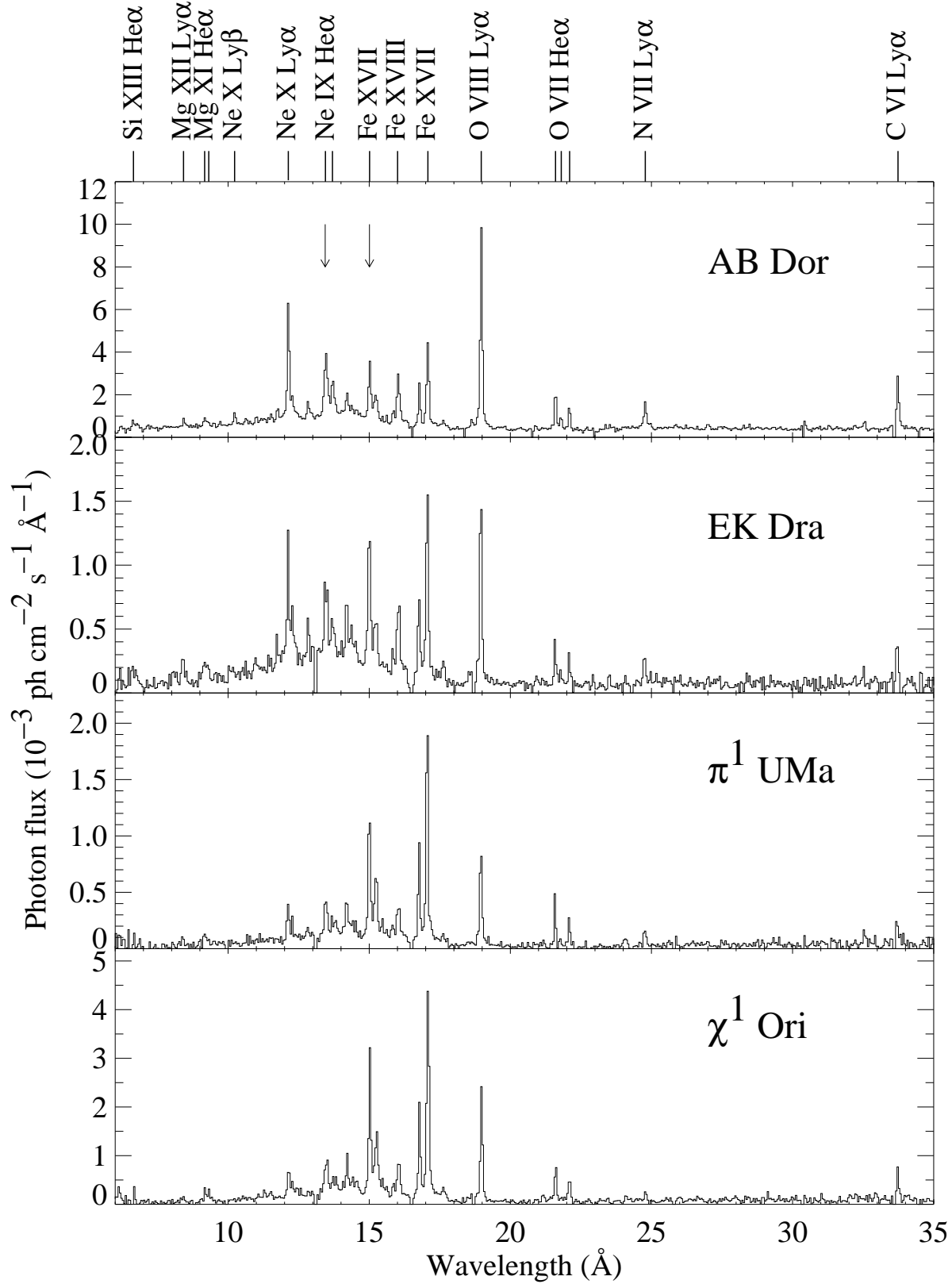}	
  \end{center}
\caption{RGS spectra of solar analogs. Their order is set similarly to
Fig.~\ref{maudard-B1_fig:fig1}. The arrows designate lines with similar
maximum formation temperature; hence different line ratios of the
{\rm Fe}~\textsc{xvii} line at 15\AA\  and the {\rm Ne}~\textsc{ix} line suggest
differences in coronal abundances in each star.}  
\label{maudard-B1_fig:fig3}
\end{figure*}

\begin{figure*}[!ht]
  \begin{center}
    \includegraphics{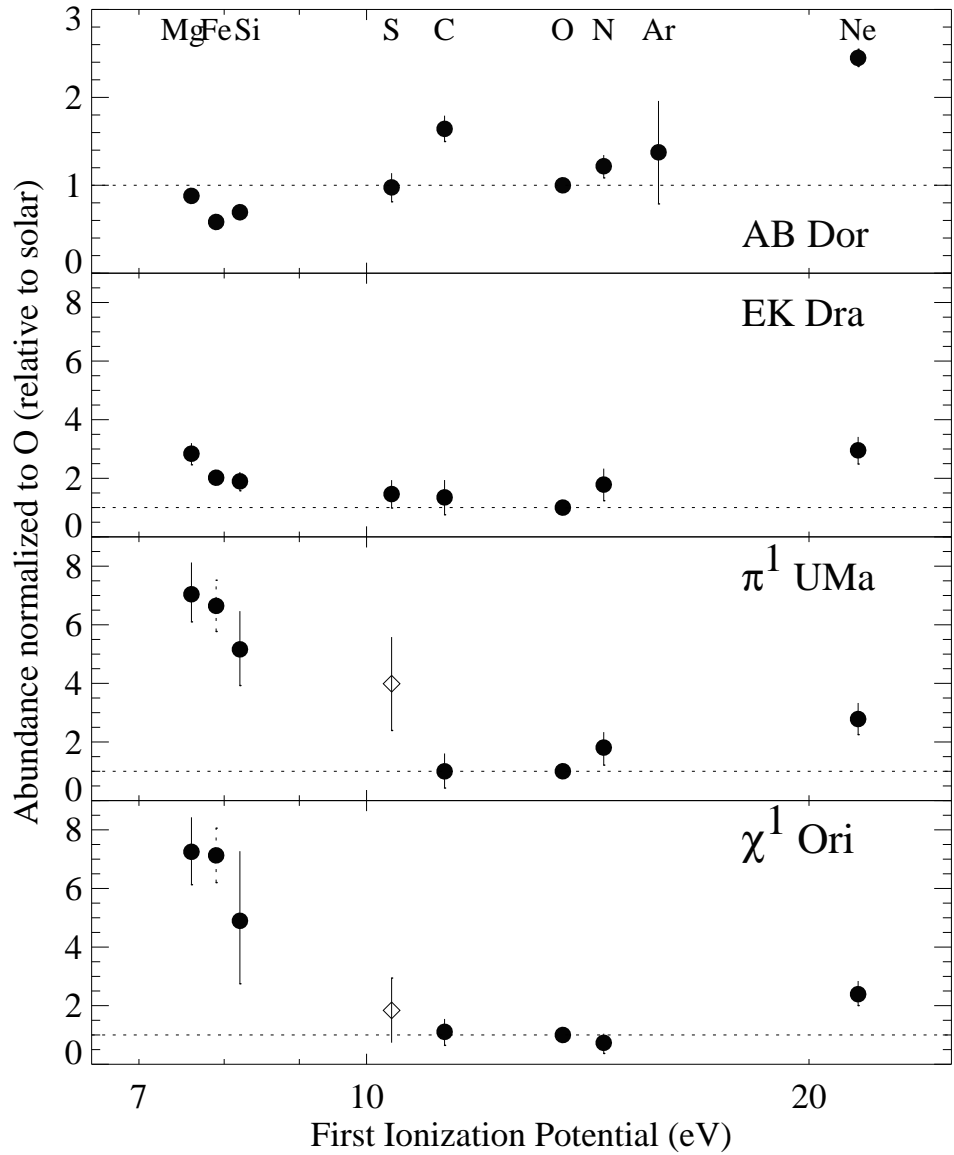}	
  \end{center}
\caption{Coronal abundance normalized to oxygen in solar analogs
as a function of the First Ionization Potential. Note that the activity level
decreases (their age increases) from top to bottom. Similar photospheric abundances
have been taken as in Fig.~\ref{maudard-B1_fig:fig2}.}  
\label{maudard-B1_fig:fig4}
\end{figure*}

Since reliable determinations of \emph{stellar} photospheric abundances
in active stars are rare, we cannot normalize the derived coronal abundances by
their photospheric counterparts. Hence, it is possible that the FIP bias
observed in HR~1099 and UX Ari is simply a reflection of their photospheric
composition. However, it is possible to use stars with known photospheric 
abundances to determine whether a FIP effect (or its inverse) exists in active
stars. We have used spectra from solar analogs (Fig.~\ref{maudard-B1_fig:fig3}) for 
which the photospheric composition is close to solar. These stars 
represent the Sun in its past and probe its activity in its infancy.
While AB Dor is strictly speaking not a solar analog (spectral
type K0), its activity is similar to that of a very active
young Sun. Similarly to the RS CVn binary systems, the differences in the
spectral features do not depend solely on the emission measure distribution, but
also on intrinsic variations of coronal abundances. Spectral fits confirm that
the coronal composition of our sample is different for each star.
Figure~\ref{maudard-B1_fig:fig4} shows the coronal abundances as a function of
the FIP, like in Fig.~\ref{maudard-B1_fig:fig2}. The FIP bias appears to
correlate with the activity level (or age): an IFIP effect is found in the most
active star AB Dor, like in the most active RS CVn binaries, while the
intermediately active EK Dra shows no pronounced bias. Finally, the oldest, less active stars $\pi^1$ UMa and $\chi^1$ Ori
display a similar abundance bias, close to that observed in the Sun. Note that the
transition in the FIP bias occurs for low-FIP elements while the high-FIP
elements appear to have similar coronal abundances (normalized to oxygen) in all
stars.

\begin{figure*}[!ht]
  \begin{center}
    \includegraphics[width=\textwidth]{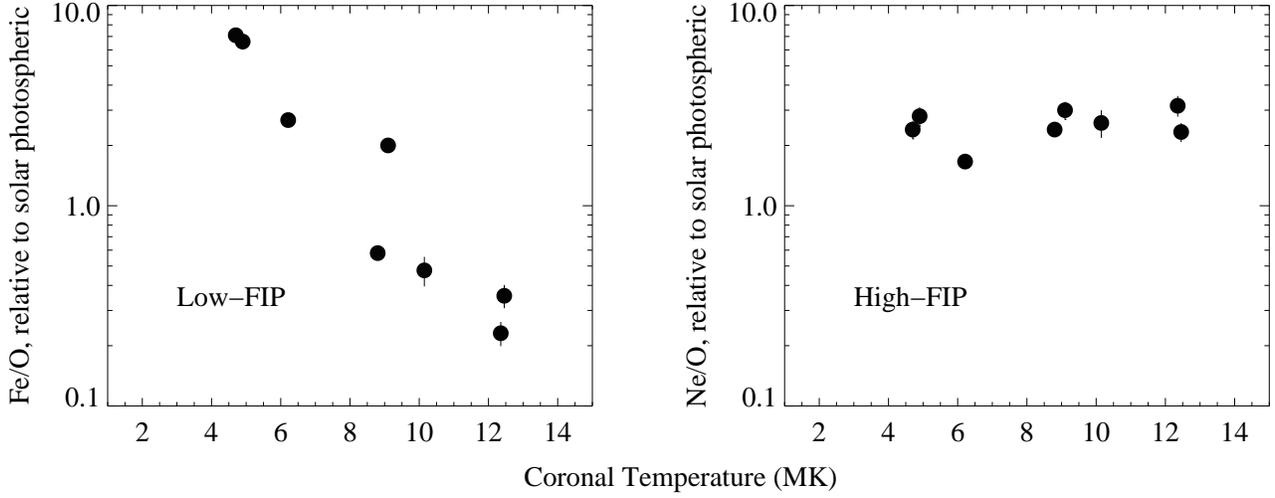}	
  \end{center}
\caption{Coronal abundances (normalized to {\rm O}) as a function of the average 
coronal temperatures, for {\rm Fe} (low-FIP; left) and {\rm Ne} (high-FIP; right). The 
data include solar analogs and RS CVn binaries.
}  
\label{maudard-B1_fig:fig5}
\end{figure*}

Based on the emission measure distributions obtained from multi-temperature fits, we
have calculated a logarithmic average temperature and defined the latter as the
average coronal temperature. While most inactive stars have temperatures
between 4 and 6~MK, RS CVn binary systems display average quiescent temperatures 
around 10--20~MK. Previous analysis showed a correlation between the
X-ray luminosity of solar analogs and their average coronal temperatures
(\cite{maudard-B1:guedel97}). The average coronal temperature can therefore be 
considered to be an activity indicator. The abundances of low-FIP elements correlate with
the coronal temperature. While they are depleted in the most active stars, a
transition occurs with decreasing temperature, and their abundances drastically
increase (relative to high-FIP elements). On the other side, the abundances of
high-FIP elements stay constant. In Figure~\ref{maudard-B1_fig:fig5} we give
examples for Fe and Ne, representing low-FIP elements and high-FIP elements,
respectively. Note that this behavior is reproduced in other low-FIP elements
(e.g., Mg, Si) and high-FIP elements (e.g., C, N). 

\subsection{Coronal abundance during flares}

\begin{figure*}[!ht]
  \begin{center}
    \includegraphics[width=\textwidth]{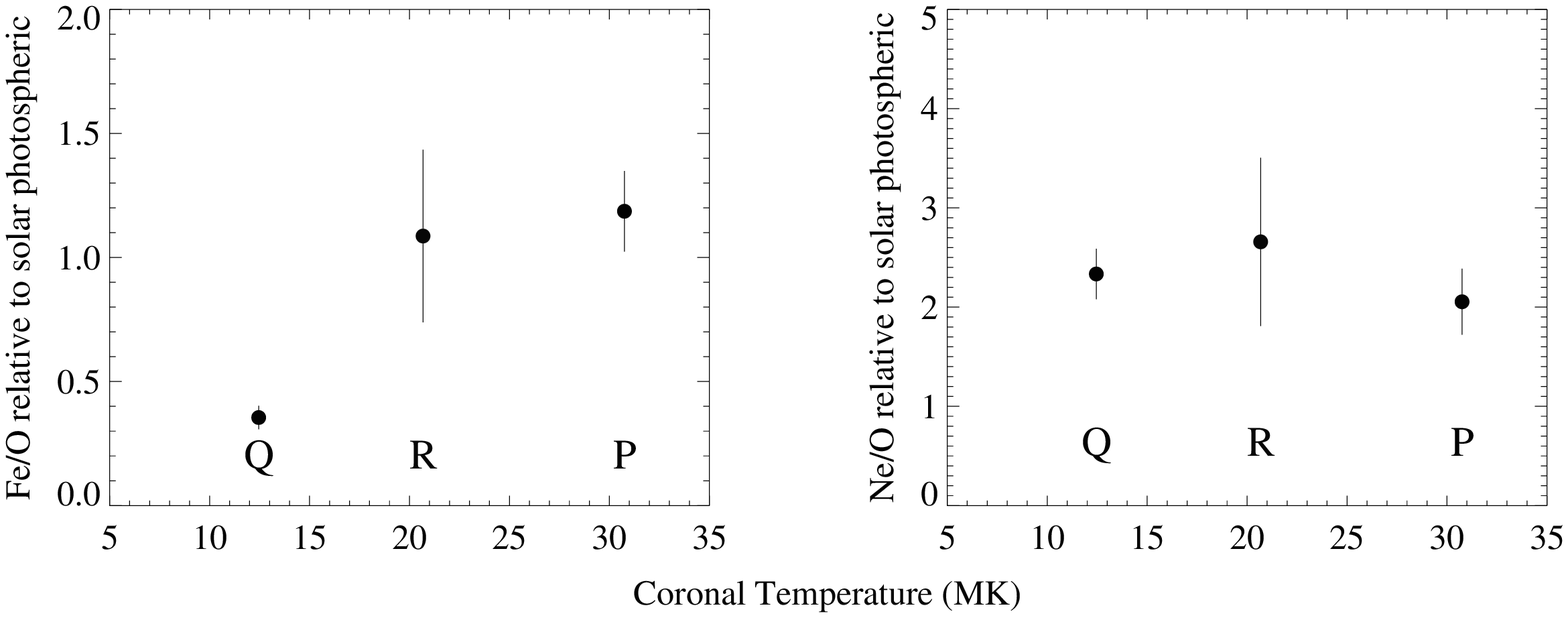}	
  \end{center}
\caption{Coronal abundances (normalized to {\rm O}) as a function of the average
coronal temperature during a large flare in HR~1099. Left panel gives {\rm Fe/O} 
ratios, while the right panel gives {\rm Ne/O} ratios. `Q' stands for quiescent, 
`R' for flare rise, and `P' for flare peak.}  
\label{maudard-B1_fig:fig6}
\end{figure*}

In the previous section, coronal abundances of active stars in \emph{quiescence}
have been reported to show a transition from under- to overabundant low-FIP
elements with decreasing activity (or coronal temperature), while the high-FIP
elemental abundances stay constant (relative to oxygen). However, previous data showed
that the average metallicity Z or the Fe abundance can increase during large
flares (e.g., \cite{maudard-B1:ottmann96}). \cite*{maudard-B1:guedel99}
obtained time-dependent measurements of several elemental abundances
during a large flare in UX Ari with \textit{ASCA}. The abundance of low-FIP elements 
increased more significantly than those of high-FIP elements. Recently,
\cite*{maudard-B1:audard01b} found a similar behavior in the \textit{XMM-Newton} 
data of a flare in HR 1099. We have redone the analysis of this flare using a more
recent calibration. Figure~\ref{maudard-B1_fig:fig6} shows the Fe/O and Ne/O
ratios versus the coronal temperature of HR~1099, before the flare (quiescence),
during the flare rise, and at flare peak (no complete decay available).
Consistently with our previous analysis (\cite{maudard-B1:audard01b}), we have
found that the absolute Fe abundance increases during the rising part of the
flare. In contrast, the absolute Ne abundance stays constant.  
Other low-FIP elements and high-FIP elements show similar respective behavior.
Note, however, that the signal-to-noise ratio in the RGS time-dependent
spectra did not allow us to better sample the flare event. Longlasting strong
flares are needed to obtain high-quality spectra of a flare. 

\section{Conclusions}
High-resolution X-ray spectra of magnetically active stars have been
investigated with the Reflection Grating Spectrometers on board
\textit{XMM-Newton}. The high-energy data ($> 1.5$~keV) of the EPIC CCD spectra
were used to better constrain the high-temperature part of the emission measure
distributions and to profit from the presence of H-like and He-like transitions
of Si, S, Ar, and Ca. It was found that the most active stars, such as the bright
RS CVn binary systems, show a marked depletion of low-FIP elements (e.g., Fe, Mg,
Si) relative to high-FIP elements (e.g., C, N, O, Ne), opposite to the 
FIP effect observed in the solar corona. This ``inverse FIP'' effect is however 
not observed in the intermediately active RS CVn binary Capella. Since their 
photospheric abundances are mostly unknown or not reliable, one can hypothesize 
that the observed FIP bias is simply a reflection of their photospheric composition.

To remove the uncertainty of surface abundances, we have analyzed
high-resolution X-ray spectra of solar ana\-logs of known photospheric composition
(close to solar). These solar-like stars span a wider range of coronal activity
(from inactive to active) and represent the evolution of the solar corona in
time. We have found a transition from a depletion of low-FIP elements (relative
to high-FIP elements) in the most active stars toward a marked enhancement of
their abundances in the inactive stars. On the other hand, the abundances of
high-FIP elements do not vary with the activity level (or coronal
temperature), relative to O. The IFIP effect found in the active RS CVn binary systems fit well
into this transition, under the assumption that their photospheric composition
is also close to solar. Similarly, the solar FIP effect (enhancement of low-FIP
elements by factors of 4--8) fits into this picture. However,
although the scenario of correlating the activity level with the FIP bias is
tempting, it may be too simplistic; indeed, such scenario does not explain the
absence of any FIP bias in the corona of the old, inactive Procyon
(\cite{maudard-B1:drake95}).

We have put forward first ideas to explain the inverse FIP effect seen in active 
stars: downward propagating electrons detected by their gyrosynchrotron emission
in active stars could prevent chromospheric ions (mostly low-FIP elements) from escaping into 
the corona by building up a downward-pointing electric field
(\cite{maudard-B1:guedel02}). As the density of high-energy electrons 
decreases with decreasing activity, the inverse FIP effect is quenched. During 
large flares, however, the high-energy electrons heat a significant portion of 
the chromosphere to bring up a near-photospheric mixture of elements into the
corona, and this  effect has indeed been observed
(\cite{maudard-B1:guedel99,maudard-B1:audard01b}). 
The new results by \textit{XMM-Newton} and \textit{Chandra} have opened a 
new field of research relevant to the physics of heating and dynamics of outer 
stellar atmospheres.

\begin{acknowledgements}

We acknowledge support from the Swiss National 
Science Foundation (grant 2100-049343). This work is based on observations 
obtained with XMM-Newton, an ESA science 
 mission with instruments and contributions directly funded by ESA Member 
 States and the USA (NASA).

\end{acknowledgements}

\end{document}